\pdfoutput=1
\documentclass[sigconf]{acmart}

\AtBeginDocument{%
  \providecommand\BibTeX{{%
    \normalfont B\kern-0.5em{\scshape i\kern-0.25em b}\kern-0.8em\TeX}}}

\setcopyright{acmcopyright}
\copyrightyear{2018}
\acmYear{2018}
\acmDOI{XXXXXXX.XXXXXXX}

\acmConference[Conference acronym 'XX]{Make sure to enter the correct
  conference title from your rights confirmation emai}{June 03--05,
  2018}{Woodstock, NY}
\acmBooktitle{Woodstock '18: ACM Symposium on Neural Gaze Detection,
 June 03--05, 2018, Woodstock, NY} 
\acmPrice{15.00}
\acmISBN{978-1-4503-XXXX-X/18/06}

\usepackage{multicol}
\usepackage{multirow}
\begin{document}

\title{ChatCoder: Chat-based Refine Requirement Improves LLMs' Code Generation}


\author{Zejun Wang}
\affiliation{%
  \institution{Key Lab of HCST (PKU), MOE; SCS}
  \city{Beijing}
  \country{China}
}

\author{Jia Li}
\affiliation{%
  \institution{Key Lab of HCST (PKU), MOE; SCS}
  \city{Beijing}
  \country{China}
}

\author{Ge Li}
\affiliation{%
  \institution{Key Lab of HCST (PKU), MOE; SCS}
  \city{Beijing}
  \country{China}
}

\author{Zhi Jin}
\affiliation{%
  \institution{Key Lab of HCST (PKU), MOE; SCS}
  \city{Beijing}
  \country{China}
}

\renewcommand{\shortauthors}{}

\begin{abstract}
Large language models have shown good performances in generating code to meet human requirements. However, human requirements expressed in natural languages can be vague, incomplete, and ambiguous, leading large language models to misunderstand human requirements and make mistakes. Worse, it is difficult for a human user to refine the requirement. To help human users refine their requirements and improve large language models' code generation performances, we propose ChatCoder: a method to refine the requirements via chatting with large language models. We design a chat scheme in which the large language models will guide the human users to refine their expression of requirements to be more precise, unambiguous, and complete than before. Experiments show that ChatCoder has improved existing large language models' performance by a large margin. Besides, ChatCoder has the advantage over refine-based methods and LLMs fine-tuned via human response.
\end{abstract}




\maketitle

\section{Introduction}
Large language models(LLMs) have performed well in code generation. Given human problem descriptions expressed in natural language, LLMs can generate corresponding code to meet human requirements. Not only do the well-known close-source LLMs for business show the ability to generate code with high quality (e.g., GPT-4\cite{gpt-4} pass 67\% of the tests in HumanEval\cite{codex_humaneval}), but also the recent open-source LLMs have reported their good results on code generation (e.g., Gunasekar et al. have designed an open-source LLM called phi-1\cite{textbook_all_you_need} which has passed 50.6\% of the tests in HumanEval). Thus, applying LLMs to assist human programmers in their everyday coding tasks is promising.

\begin{figure*}[htbp]
    \centering
    \includegraphics[scale=0.6]{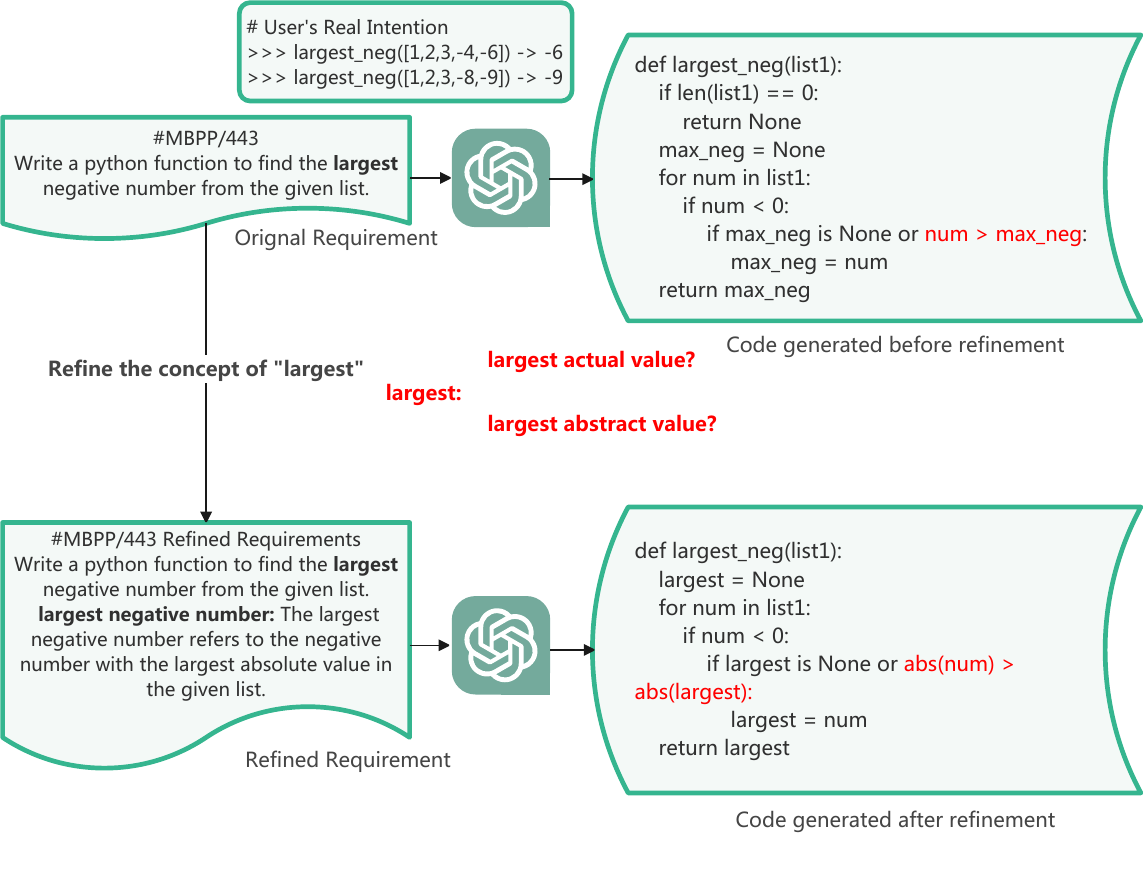}
    \caption{Example of Refinement Improving Code Generation Performance}
    \label{fig:example_0}
\end{figure*}

However, human's poor requirement expressions in natural language restrict LLMs' ability to generate better programs. Human expressions can be vague, incomplete, and ambiguous. These low-quality requirement expressions mislead large language models to generate the wrong code. We raise an example from the sanitized-MBPP dataset\cite{mbpp} in Figure \ref{fig:example_0} to illustrate the issue, which is thought unambiguous by the authors. Suppose that we want gpt-3.5-turbo to write a function to find the largest negative number from the given list. Based on the original requirement, the large language model generates a program which can extract the negative numbers with the largest \textbf{actual} value correctly. However, the authors of sanitized-MBPP think that the 'largest negative number' means the largest \textbf{absolute} value. Thus the large language model generates the wrong code due to the bad expression 'largest'.

% 这一段和下面一段可以融合起来：
% - 上述问题可以用需求精化来解决；
% - 需求精化是什么？
% - 在本文中，我们使用需求精化来提升人和LLMs之间的交互...
The problem can be solved via requirements refinement. Requirements refinement is the process of revealing the underlying dependencies and hidden structures\cite{re_framework}. With more details revealed, incomplete information will be filled up during requirement refinements, and the ambiguities will be clarified. In our example illustrated in Figure \ref{fig:example_0}, we can simply reveal the hidden structure of 'the largest' as 'the largest absolute value' to the large language model. With the refined requirement, the large language model generated the code that fulfilled the MBPP's authors' expectations.

Requirements refinement asks for the collaboration of human users and large language models. In the context of requirement engineering, requirements refinement is performed by a series of interactions between the software supplier (the coder) and the software customer (the user). The software supplier analyzes the initial expression of the customer's requirements and raises the points of refinement. The software customers need to respond to the points based on which the supplier can finish a round of refinement. Neither the software customer nor the software supplier is qualified to perform requirements refinement by themselves. According to IEEE Std 830-1998\cite{ieee_re}, customers usually do not understand the software design and development process well enough to write a usable one. Suppliers usually do not understand the customer's problem and field of endeavour well enough to specify requirements for a satisfactory system. In the scenario of asking LLMs to generate programs to fulfil human requirements, the human user of LLM is the customer, and the LLM itself is the supplier. To let the supplier LLM produce code that better fulfils the user's requirements via requirements refinement, we need to develop a method for humans and LLMs to collaborate.

%% No need for the code. just show that the LLM gets confused. 
%% shrink the size and put it to the first page (single-column)

% 本文提出ChatCoder，一个新的代码生成方法
% 核心创新是：提出了一种chat 方式来实现人和大模型之间的需求精化，然后提升代码生成的准确率
% 整体流程
We propose ChatCoder, a new method for code generation with large language models through requirements refinement via chat. It is a concise dialogue framework that assists LLMs and humans' collaboration on requirements refinement via chatting. The key problem is how to chat with the large language model. Our solution, ChatCoder, has a novel chatting schema designed inspired by \textit{IEEE Recommended Practice for Software Requirements Specifications}(IEEE SRS)\cite{ieee_re}. This paper mainly discusses method-level code generation. Referring to the contents of software requirement specifications raised by IEEE SRS covering every corner of a software's life cycle, we raise six angles covering the development of a method and provide the angles to large language models to analyze the requirement specifications. Then the large language models lead the human user to refine the requirements based on its analysis by adding information, correcting mistakes, giving examples, and answering questions. The whole process is in the chat form. In this paper, we test ChatCoder on the HumanEval dataset and Sanitized-MBPP dataset, and the test results show that the refined requirements with ChatCoder improve the LLM's code generation performances by a large margin, at an average of the percentage of 10. The results show that ChatCoder's refinement is effective and efficient.

% 在上述的流程中，如何chat是一个关键问题。
% 基于IEEE SRS设计了一种新颖的schema...
% 多个角度...

% 实验结果
% 数据集、指标
% 关键性的结果（比较的提升）

Our contribution is summarized as follows:

\begin{itemize}
    \item We find and raise the problem that human's poor requirement expressions in natural language limit LLMs' ability to generate better programs.
    \item We point out the necessity to ask for the collaboration of humans and large language models.
    \item We raise ChatCoder, a dialogue framework effectively assisting human and LLM's collaboration on requirements refinement for better code generation.
\end{itemize}

\section{background}
\subsection{Large Language Model for Code Generation}
Large language models are currently pre-trained Transformer-based language models with at least tens of billions of parameters. The first well-known large language model is GPT-3 \cite{gpt-3} proposed by OpenAI, and GPT-3 presented its extraordinary code generation ability. Following GPT-3, a series of business-oriented close-source large language models have been proposed, e.g. GPT-3.5 and GPT-4, whose code generation abilities improve day by day. Besides, several open-source large language models for code-related tasks have been published, e.g., StarCoder\cite{star_coder}, CodeT5+\cite{codet5+}. WizardCoder\cite{wizard_coder} etc. They have been proven to have comparable code generation capabilities with the close-source large language models.

%% correct the term usage
% prompting techniques, Chain-of-Thought Prompting
The current way of applying large language models in code generation is via prompting techniques. A prompt is a formatted text wrapping the user's original instruction for the large language model. Then the prompt is sent to the large language model as input to get the large language model's response. Given the user's description of a programming task, properly designed prompts will make it easier for large language models to generate the correct corresponding code. For example, Li et al. \cite{li}propose that providing examples closely related to the programming tasks can help large language models to generate better code. Jiang et al. \cite{jiang}propose that appending the text that encourages the LLMs to decompose the programming task helps large language models solve complicated problems. In this paper, our proposed method can be categorized as prompt engineering as well.

\subsection{Requirements Refinement}
Requirements refinement is both a process of deriving specifications and a necessary means towards preparing architecture designs. During requirements refinement, the design of requirement specifications should reveal its underlying dependencies and hidden structure. Requirements refinement is the start from requirements to implementation design. It is important because many users in practice do not understand what functions they want precisely at the beginning of a software project\cite{sofl_analyse}. With requirements refinement, the users and software suppliers can agree on what function the user truly needs.

Previous studies of requirements refinement focus on providing a formal method for the software supplier to analyse and refine the software customer's requirements. Liu \cite{re_framework} raises a hierarchical framework from the business level to the component level to refactor the customer's requirements. Darimont and Lamsweerde \cite{kaos_analyse} propose formal refinement patterns for goal-driven requirements elaboration via KAOS. Liu\cite{sofl_analyse} proposes to use the SOFL language to describe the refinement process and raises the model of successive refinements in which the requirements refinement is a process from coarse to fine with a loop back. Jong et al. propose to use nondeterminism and parameterised specifications to support step-wise specifications and have the specifications written and analysed using the language and proof checker of PVS.

Requirements refinement requires collaboration with both the software provider and the software user. On the one hand, requirement analysing methods, e.g., the formal approaches mentioned above, can not achieve complete and accurate themselves. Domain knowledge and the customer's personal quality (e.g., the ability to express oneself clearly) are essential in requirements refinement. On the other hand, the users can not refine their requirements by themselves since they may not understand the software design enough, leading to the phenomenon that their proposed requirements may not fulfil what they actually want. \textbf{So what is important is to find a friendly way of interaction:} the software provider find the points in the user's requirement which need refinement, then the provider asks for the software user's comments on the refinement in the way that the software user can understand and give proper answers.

\section{Methodology}
\subsection{Overall Structure of ChatCoder}
%% edit: reformat into code generation framework
ChatCoder is code generation method through requirements refinement via a dialogue framework designed for the communication between a large language model and its user to refine the requirements. Within the framework, a large language model can analyse the arguments to refine the user's original requirement expression, then return the arguments back to the users in a way that human users can easily understand and give responses.

\begin{figure*}[htbp]
    \centering
    \includegraphics[scale=0.6]{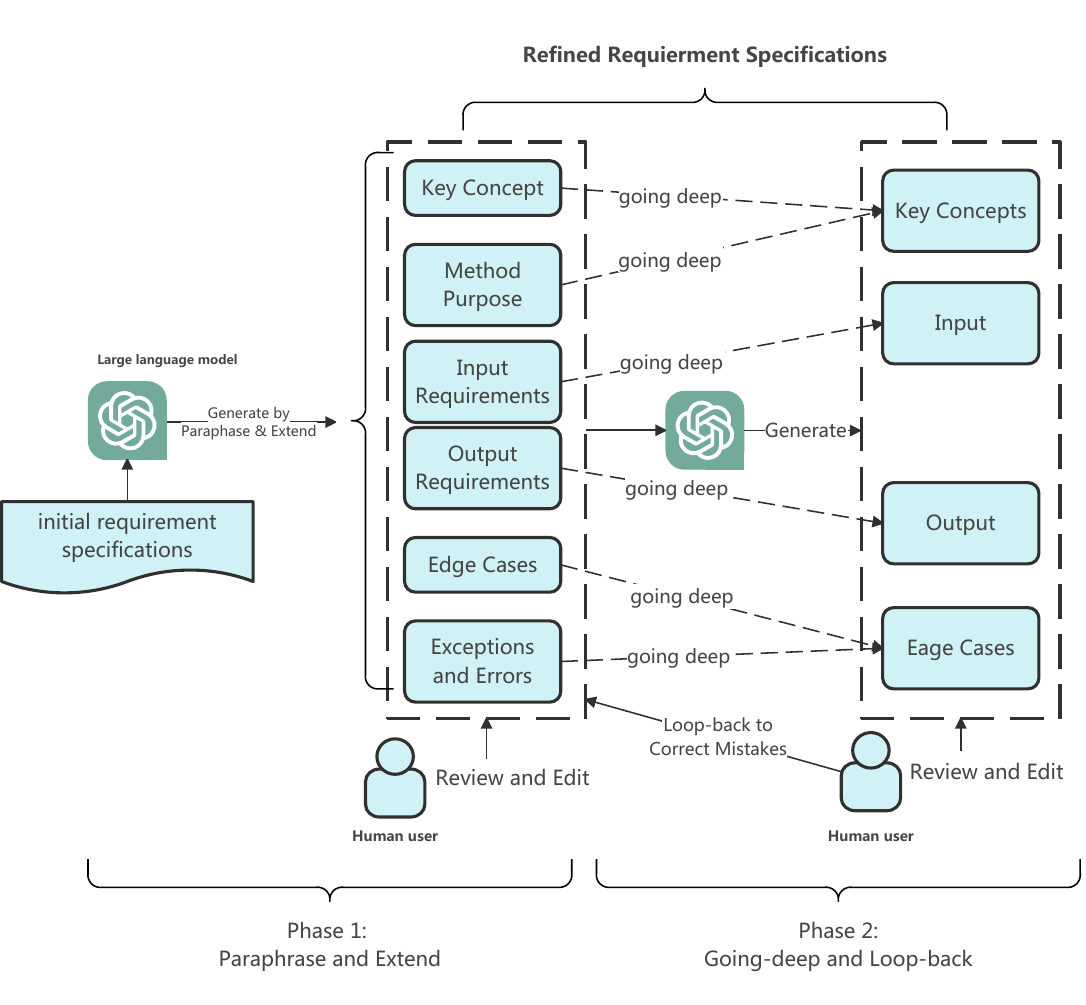}
    \caption{Overall Structure of the ChatCoder Dialogue Framework}
    \label{fig:method}
\end{figure*}

The overall structure of ChatCoder is a two-round dialogue illustrated in Fig \ref{fig:method}. The first round is \textit{Paraphrase and Extend}. Since the human user's expression of requirements can be vague, incomplete and ambitious, ChatCoder uses prompts to ask the large language model to paraphrase the user's original requirements from several angles that complete requirement specifications must be clear. For the missing or ambitious arguments which require refinement, ChatCoder asks the large language model to extend them with its assumptions gained from its training data. Human users need to review the refined specifications and correct the mistakes within. The second round is \textit{Going-deep and Loop-back}. In this round, ChatCoder requires large language models to ask the human users about their confusion about the refined specifications in \textit{Paraphrase and Extend} for losing information and further refinement. Human users need to answer the questions and loop back to correct the refined specifications when the users find the large language model's questions are raised based on wrong requirement specifications. After the two rounds of refinement, the refined requirement is obtained and then sent to large language models to get the user's desired programs.

In the following paragraphs, we will explain the design of each round in detail.

\subsection{Paraphrase and Extend}
%% Requirement 慎用
The large language model is asked to paraphrase the user's initial requirement expression in this round. The paraphrase is performed by extending the initial requirement on the preset angles extracted from existing research of requirement engineering. ChatCoder wraps the instruction to paraphrase the user's requirement and the angles used for extension in a prompt to order the large language model to perform the paraphrase. Then the prompt is sent to the large language model for its response. The format of the prompt is presented in Figure \ref{fig:prompts}.

The angles selected are based on the environment of applying ChatCoder. Since this paper mainly discusses generating method-level programs, the angles for ChatCoder are all about method-level requirements refinement. In particular, the ChatCoder in this paper has five angles for the \textit{Paraphrase and Extend} round, inspired by \textit{IEEE Recommended Practice for Software Requirements Specifications}:

\begin{itemize}
    \item \textbf{Key Concepts} This angle asks the large language model to extract and explain the key concepts involved in the user's requirements, including objects and actions. By extending this angle, the user and the large language model can align their understanding of the key concepts, setting a firm basis for further discussion.
    \item \textbf{Method Purpose} This angle asks the large language model to paraphrase the function provided by the method to be implemented. In this angle, the large language model will describe the transformation for the input and the changes of the running states in a more detailed way. The LLM's description reflects its ongoing implementation based on the LLM's understanding of the initial requirement expression and its inference for the incomplete expression, revealing the error and incompleteness of the requirement expression.
    \item \textbf{Input Requirements} This angle asks the large language model to extend the requirements for the method's inputs, including the parameters' types, actual meaning, boundaries and properties. Explaining the meanings is another chance for the LLM and the user to align their understanding of the requirements. The type, boundary and property are easily missing but play important roles in the design of the algorithm.
    \item \textbf{Output Requirements} This angle asks the large language model to extend the requirements for the method outputs, including the types, the meaning and the format. Explaining the meanings is another chance for the LLM and the user to align their understanding of the requirements. While a method may serve other methods, its returning type and format matter but can be missing, e.g., the decimals to reserve for a floating-point output number.
    \item \textbf{Edge Cases} This angle asks the large language model to extend possible edge cases and solutions. Since a method can run in complicated outer environments, the input and the global variable states may not fulfil the method's running preconditions. So properly handling edge cases is necessary for a robust method implementation but can be easily ignored by software customers. 
    \item \textbf{Exceptions and Errors} This angle asks the large language model to extend the solutions for possible exceptions and errors during the method's execution. Like edge cases, handling exceptions and errors are necessary but can be easily missed by the users because of their unprofessional software design. The large language model must analyse, raise solutions and wait for the users' review.
\end{itemize}

The human user is supposed to review the large language model's response to the instructions for refining requirements. For the key concepts and method purpose, the human user is requested to correct the mistakes made by the large language model. For the input and output requirements, the human user is requested to correct the mistakes for the meanings and review whether the large language model's inference on the input and output formats meets the real needs. For the edge cases, exceptions and errors, the users are requested to review whether they can occur and whether the large language model's proposed solution is satisfactory. If the human user encounters an expression that is difficult to understand and rewrite, the user can directly delete the expression.

Our design of \textit{Paraphrase and Extend} is an effective and efficient way for the large language model and the human user to communicate for requirements refinement. First, our instructions for the large language model are in natural language. Compared with the formal language designed for human coders to analyse the requirements for refinement, large language models are more familiar with natural languages since most of their training data is in natural languages. Second, the angles mentioned in the instructions cover many reasons humans and AI programmers make mistakes. Refining the requirements from these angles helps reduce programming mistakes. Third, it is easy for human users to read, understand and modify the refined requirements, thanks to the LLM's string expression power. Most of the refined specifications are generated by the large language model. All the work left for human users is only to make modifications, which is a small workload compared to generating the whole refined specifications, not to say that human users may not know what to write for the refinement.   

%% look for alternatives
\subsection{Going-deep and Loop-back}

In this round, the large language model is asked to going-deep: to further refine the requirements based on the specifications obtained in \textit{Paraphrase and Extend}; the human user is requested to loop back: check for possible errors in the reviews and the errors corrected.

\begin{figure*}[htbp]
    \centering
    \includegraphics[scale=0.8]{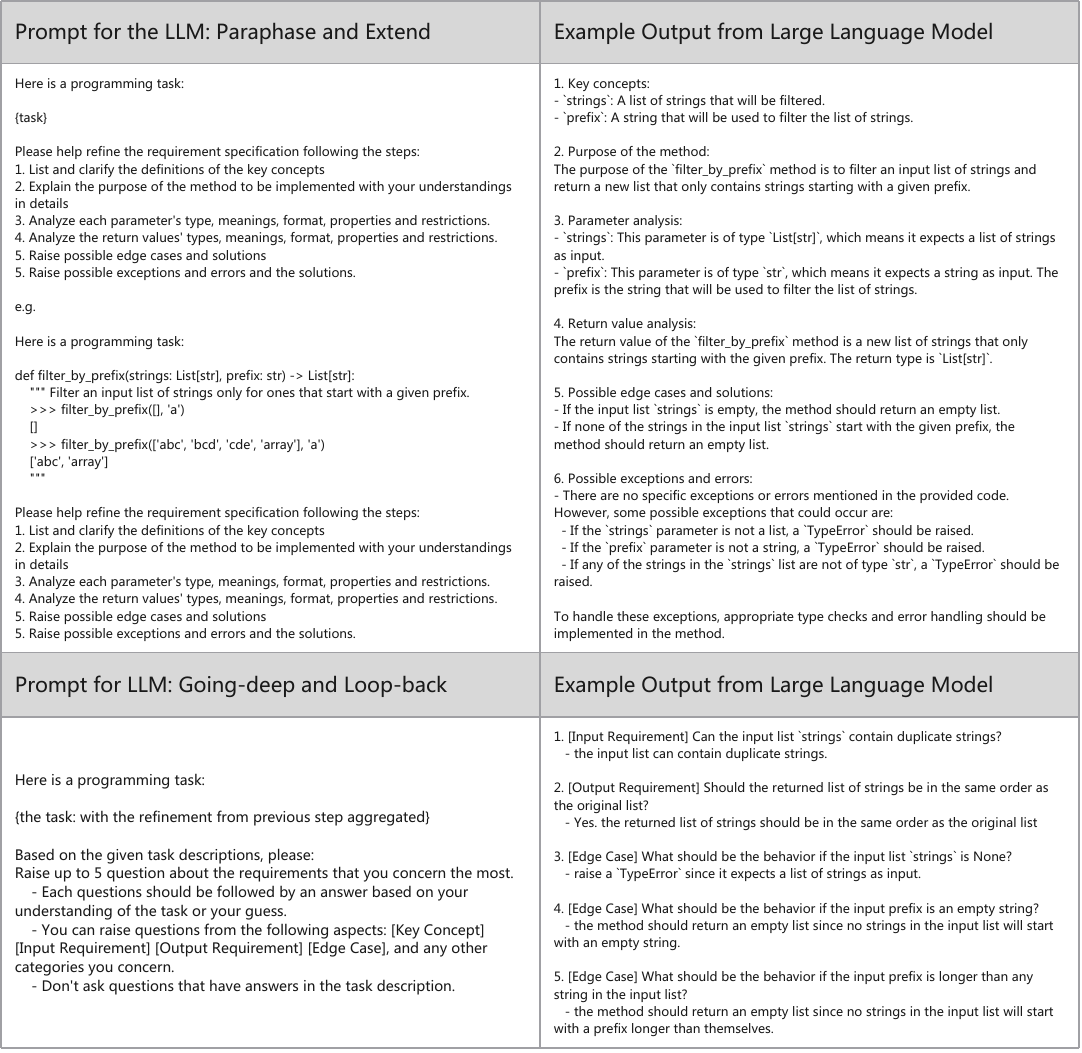}
    \caption{Prompts for Large Language Models and Example Outputs}
    \label{fig:prompts}
\end{figure*}

\textit{Going-deep} The large language model is asked to raise questions in the angles based on the existing specifications obtained in \textit{Paraphrase and Extend}. The instruction for the large language model is also wrapped in a prompt, presented in Figure \ref{fig:prompts}. We design \textit{Going-deep} to refine the requirements further because the large language model is a black box, and it is hard to say we have used up its potential to refine requirements through \textit{Paraphrase and Extend}. In this round, we let the large language model ask questions in a free form for what confused the most about the requirements, then give possible answers based on its observations or assumptions. Suppose the large language model keeps raising questions which are answered in the specifications. In that case, we regard the specifications are complete enough for the large language model to generate corresponding programs.

\textit{Loop-back} The user is asked to review the questions and answers generated by the large language model in \textit{Going-deep} and correct the wrong answers for further refinement. The user may find that the large language model raises wrong questions, e.g., it asks whether the output list should be sorted. However, the desired output is an integer. In this scenario, the user must "Loop-back": review the specifications in \textit{Paraphrase and Extend} to look for the wrong expressions leading to the wrong questions, then have them corrected. Loop-back is important because it is difficult to guarantee that the users never make mistakes. 

After \textit{Going-deep} and \textit{Loop-back}, the user will have the updated specifications from \textit{Paraphrase and Extend} and the further refined specifications from \textit{Going-deep}. Then these refined specifications are appended to the original expression of requirements and sent to the large language model to get the large language model's generated programs.

\section{Experiments}
\subsection{Experiment Settings}
\textbf{Datasets:} We select three datasets for our experiments:
\begin{itemize}
    \item \textit{Sanitized-MBPP} A well-known and widely-used dataset\cite{mbpp}. Its test set contains 257 programming questions which standalone Python methods can solve. We choose this dataset for the following reasons. First, its task descriptions are short, which means they are more likely to be incomplete and ambiguous than longer descriptions, so we can find out whether ChatCoder can let LLMs analyze the points within each description for refinement. Second, the authors of \textit{Sanitized-MBPP} claimed that these task descriptions are manually checked for disambiguation. It provided a chance to validate whether ChatCoder can make LLM analyze the task description from a different angle from the dataset's authors.
    \item \textit{HumanEval} A well-known and widely-used dataset\cite{codex_humaneval}. It has 164 programming questions to be solved by Python programs. We chose this dataset because its task descriptions are longer and more complicated than those of \textit{Sanitized-MBPP}, from which we want to evaluate whether ChatCoder can still find the points for refinement and keep improving LLMs' code generation performances.
\end{itemize}

\textbf{Baselines: } We select four baselines for our experiments:
\begin{itemize}
    \item \textbf{gpt-3.5-turbo}. The latest version of the gpt-3.5-turbo family, a family of closed-source large language models published by OpenAI. It is powerful enough and easy to access, leaving the time long enough for anyone to reproduce our experiments before it is deprecated. 
    \item \textbf{gpt-4}. The newest generation of the closed-source large language model, published by OpenAI, performs extraordinarily well on code generation.
\end{itemize}

\textbf{Generation Configurations} For \textit{HumanEval}, we perform greedy generation, which means the generation is zero-shot, and the sampling is performed only once with a temperature of 0. For \textit{Sanitized-MBPP}, we perform 3-shot generation. For each task, we sample 20 programs with top\_p=0.2 when evaluating models for gpt-3.5-turbo. As for GPT-4, because there is a calling rate limit and the calling fee is high, it is difficult and expensive to sample 20 programs for a programming task. So we sample one program for a programming task with temperature 0 like HumanEval. The version of GPT-4 is gpt-4-0613. The version of gpt-3.5-turbo is gpt-3.5-turbo-0613. For a fair comparison, we rerun all the baselines with the same prompts and our generation configuration rather than copy the results from the original papers.

\textbf{Metrics} We report the test pass rate\cite{codex_humaneval}. For HumanEval and Sanitized-MBPP on GPT-4, we report pass@1. We report pass@1, pass@2, pass@5, and pass@10 for the other settings.

\subsection{Research Questions}
To evaluate our proposed ChatCoder, we raise and investigate the following research questions:
\begin{itemize}
    \item \textbf{1)} How does ChatCoder perform compared with existing code generation models?
    \item \textbf{2)} Is ChatCoder an efficient method for LLM and human users to communicate for requirement refinement?
    \item \textbf{3)} How much improvement is brought by human involvement in ChatCoder?
\end{itemize}
\subsection{RQ1: Code Generation Performances}
RQ1 is to evaluate ChatCoder's overall code generation performances compared with the baselines. Our results are reported in Table \ref{tab:main_result}. When investigating RQ1, we try ChatCoder with GPT-4 and gpt-3.5-turbo-0613. We performed \textit{Paraphrase and Extend} and \textit{Going-deep and Loop-back} on gpt-3.5-turbo-0613 and obtained the refined requirement specifications. Then, we feed these refined requirement specifications to GPT-4 and gpt-3.5-turbo-0613 to get their generated code and test the pass rates. We obtained the refined requirement specifications from gpt-3.5-turbo-0613 because it is an LLM with the ability to perform requirement analysis and is easy to access. Compared with gpt-3.5-turbo-0613, GPT-4 is both expensive and strict with access.

\begin{table*}[htbp]
    \centering
    \caption{Code Generation Performances}
    \begin{tabular}{c c | c  c  c c}
    \toprule
     & HumanEval & \multicolumn{4}{c}{Sanitized-MBPP} \\
     \midrule
     & pass@1 & pass@1 & pass@2 & pass@5 & pass@10\\
     \midrule
     gpt-3.5-turbo & 70.12\% & 57.04\% & 58.17\% & 59.13\% & 59.75\% \\
     gpt-4 & 81.10\% & 66.15\% & - & - & - \\
     \midrule
     ChatCoder(gpt-3.5-turbo) & 79.87\% & 71.25\% & 73.23\% & 75.18\% & 76.25\% \\
     ChatCoder(gpt-4) & 90.24\% & 76.65\% & - & - \\
     \bottomrule
    \end{tabular}
    
    \label{tab:main_result}
\end{table*}

According to Table \ref{tab:main_result}, ChatCoder successfully helps large language models improve their generated program's execution accuracy through the refined requirements by a large margin. For example, for gpt-3.5-turbo, its pass@1 on Sanitized-MBPP is improved from 57.04\% to 71.25\%, and the margin is the percentage of 14. Compared horizontally, for both gpt-3.5-turbo and gpt-4, the performance improvements on Sanitized-MBPP is more prominent than those on HumanEval, which is because the task descriptions of Sanitized-MBPP are single sentences and method signatures, much more simple than the task descriptions of HumanEval. Thus the information for code generation of Sanitized-MBPP is far less sufficient than the information of HumanEval. As a result, when ChatCoder brings the refined requirement specifications full of additional information, the code generation performance on MBPP is more prominent than the improvement on HumanEval.

\begin{table*}[htbp]
    \centering
    \caption{Communication Efficiency Comparison}
    \begin{tabular}{c c | c  c  c c}
    \toprule
     & HumanEval & \multicolumn{4}{c}{Sanitized-MBPP} \\
     \midrule
     & pass@1 & pass@1 & pass@2 & pass@5 & pass@10\\
     \midrule
     gpt-3.5-turbo & 70.12\% & 56.95\% & 58.16\% & 59.48\% & 60.48\% \\
     \midrule
     Free Paraphrase & 78.05\% & 64.61\% & 65.47\% & 66.17\% & 66.68\%\\
     Free QA & 71.95\% & 66.47\% & 68.82\% & 70.91\% & 72.00\% \\
     \midrule
     ChatCoder & \textbf{79.87\%} &\textbf{ 71.25\%} & \textbf{73.23\%} & \textbf{75.18\%} & \textbf{76.25\%} \\
     \bottomrule
    \end{tabular}
    
    \label{tab:communicate}
\end{table*}

\subsection{RQ2: Communication Efficiency Evaluation}
We evaluate whether ChatCoder is an efficient way for large language models and humans to communicate for requirements refinement. The key of ChatCoder is the constraints, i.e., the angles provided for the large language models to analyse the initial expression of requirements for refinement and the instructions we designed to convey LLMs the angles. So we compare ChatCoder with two other ways of communicating with the large language model: 1) \textbf{Free Paraphrase}: We let the large language model paraphrase the original programming task without giving any angles and ask the human user to have it edited and corrected for cognition alignment; 2) \textbf{Free QA}: We let the large language model to ask human users questions about their confusion about the original programming task and collect the human users' responses. All these experiments are conducted based on gpt-3.5-turbo-0613. The results are presented on Table \ref{tab:communicate}

According to Table \ref{tab:communicate}, all three communication methods with LLMs for requirements refinement help the LLM improve its code generation results. This finding points out that any form of requirements refinement is useful and important in applying LLMs to generate code. Compared with ChatCoder, Free Paraphrase and Free QA do not instruct the LLM to perform certain kinds of refinement, leading to lower improvements. With careful inspection, we find that the additional contents generated by the LLM for requirements refinement surround our proposed analysis angles spontaneously. However, due to lacking explicit instructions, the refinement can not cover all the points covered by ChatCoder. So explicitly instructing the LLM with the angles for refining requirements is important for ChatCoder. Designing better instructions to order the LLM to refine requirements is part of our future work.

\begin{table*}[htbp]
    \centering
    \caption{Human Intervention Evaluation}
     \begin{tabular}{c c | c  c  c c}
    \toprule
     & HumanEval & \multicolumn{4}{c}{Sanitized-MBPP} \\
     \midrule
     & pass@1 & pass@1 & pass@2 & pass@5 & pass@10\\
     \midrule
     gpt-3.5-turbo & 70.12\% & 56.95\% & 58.16\% & 59.48\% & 60.48\% \\
     \midrule
     Auto-Refine & 68.90\% & 52.82\% & 54.77\% & 56.30\% & 57.12\% \\
     ChatCoder & \textbf{79.87\%} &\textbf{ 71.25\%} & \textbf{73.23\%} & \textbf{75.18\%} & \textbf{76.25\%} \\
     \bottomrule
    \end{tabular}
    
    \label{tab:labor_eval}
\end{table*}

\subsection{RQ3: Human Intervention Evaluation}
We evaluate how important human intervention is to ChatCoder. This experiment is to prove the argument that requirements refinement should involve the participation of both software provider and software supplier, in this paper, the human user and the large language model.

We evaluate the human intervention by comparing it with asking the large language model to paraphrase and generate further questions without human's edit and correction, referred to as 'Auto-Refine' in the following description. We compare the LLM's code generation performances of Auto-Refine and our ChatCoder. All experiments are conducted on gpt-3.5-turbo-0613. The results are presented in Table \ref{tab:labor_eval}

It is not surprising that Self-Refine hurts the LLM's code generation performances. Since ChatCoder utilizes requirements refinement to improve the large language model's code generation performance, human intervention is necessary and can not be neglected. The process of ChatCoder is to reveal the inner structure of the requirements from the given angles, which are not expressed explicitly, even with ambiguity. The answer to solving the ambiguity is known only by the human user. But Auto-Refine just guesses an answer based on the large language model's training data, representing how most people understand the requirement. Suppose the large language model's guess or explanation of the requirement is wrong without human edits. In that case, the large language model will generate code following the wrong understanding of requirements and give up the other possible understandings. Thus, Auto-Refine hurts the LLM's code generation performances.

\section{Discussion}
\subsection{Case Study}
\begin{figure*}[htbp]
    \centering
    \includegraphics[scale=0.7]{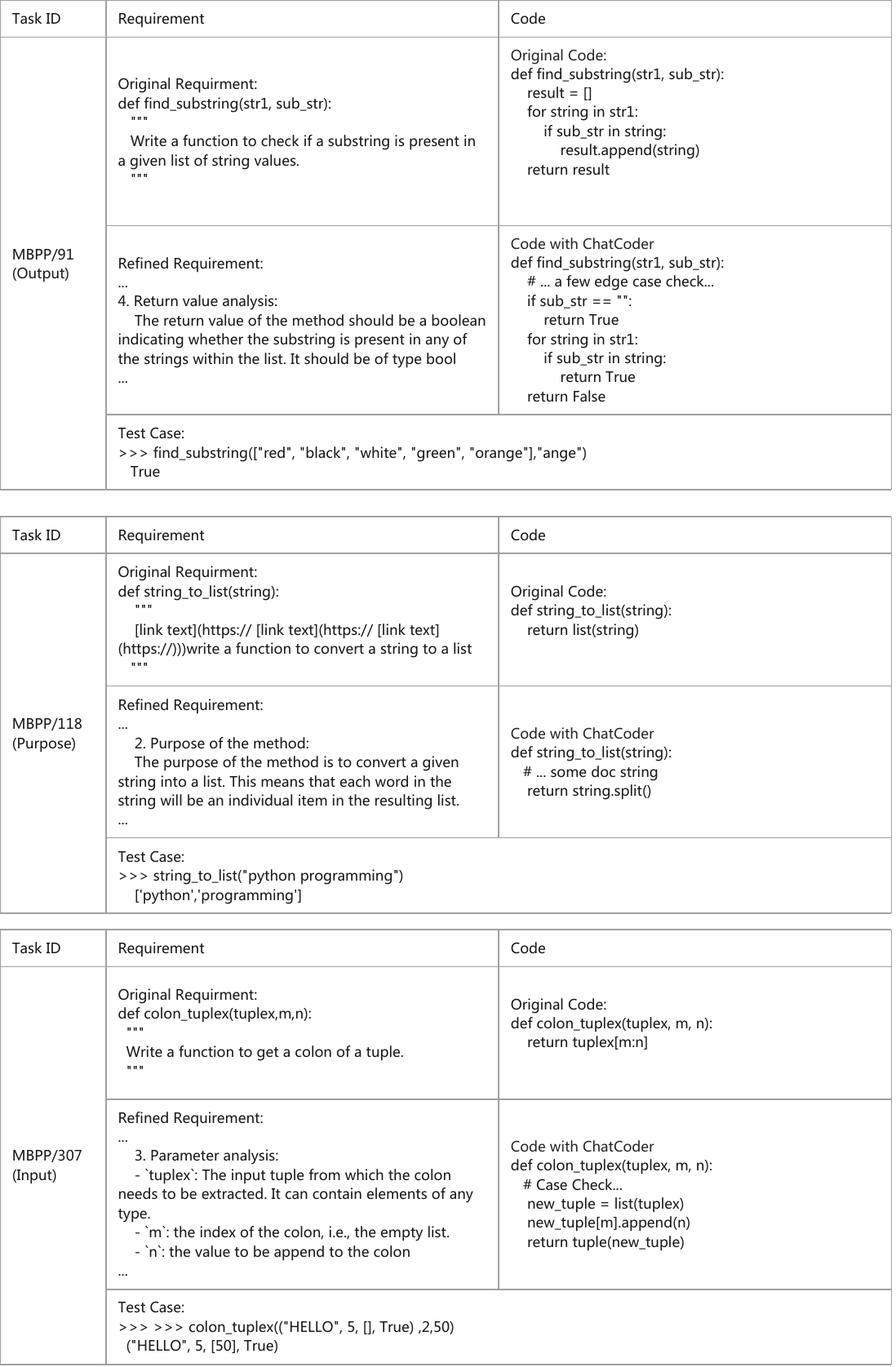}
    \caption{Case Study}
    \label{fig:cases}
\end{figure*}

This section raises several real test cases illustrating how ChatCoder helps LLMs generate code with refined requirements. Due to the page limit, we select three cases from MBPP covering refinement about the input, the output and the purpose since they influence the functional requirements directly. In contrast, edge cases and exceptions influence the robustness, requiring more space to illustrate. We put the cases in Figure \ref{fig:cases}.

\begin{itemize}
    \item \textit{MBPP/91} This task asks the coder to write a method checking if a string is presented in any string as a substring within a list. Due to the word 'if', we know the output of this method should be of judgement. However, the large language model misunderstands the task and returns a list of words. Because ChatCoder asks the large language model to analyze the output, the output requirement is refined, indicating that the method should return a boolean value. The large language model generates the correct code based on the refined requirement.
    \item \textit{MBPP/118} This task asks the coder to write a method converting a string to a list. The large language model misunderstands the purpose of the method. The string should be split into words. However, the LLM thinks the method should be split into characters. The purpose of this method expressed by the original requirement is incomplete. ChatCoder asks the LLM to analyze the purpose of the method, and the LLM returns that the method should split the string by characters. The human user reviews the refined expression and corrects this mistake. With the corrected refined requirement, the large language generates the correct code.
    \item \textit{MBPP/307} This task asks the coder to write a method to get a colon of a tuple. However, the expression is incomplete. The meaning of the parameters is missing, requiring refinement. Without refinement, the large language model thinks 'm' and 'n' are some indexes, leading to generating the wrong code. ChatCoder asks the LLM to analyze the meaning of each input parameter. The LLM responds that 'm' and 'n' are the index of the colon, which is wrong. The human user reviews the refined specification and corrects the meaning that 'm' is the index of the colon and 'n' is the value to be appended to the colon. The large language model generates the correct code with the corrected refined requirement.
\end{itemize}
\subsection{Savings of Human Labor Costs}

Compared with performing requirements refinement with requirement engineers, ChatCoder asks the large language model to generate most of the text. At the same time, human users just need to review and edit, saving lots of human labour. This section will analyze how much human labour costs are saved.

\begin{figure}[htbp]
    \centering
    \includegraphics[scale=0.4]{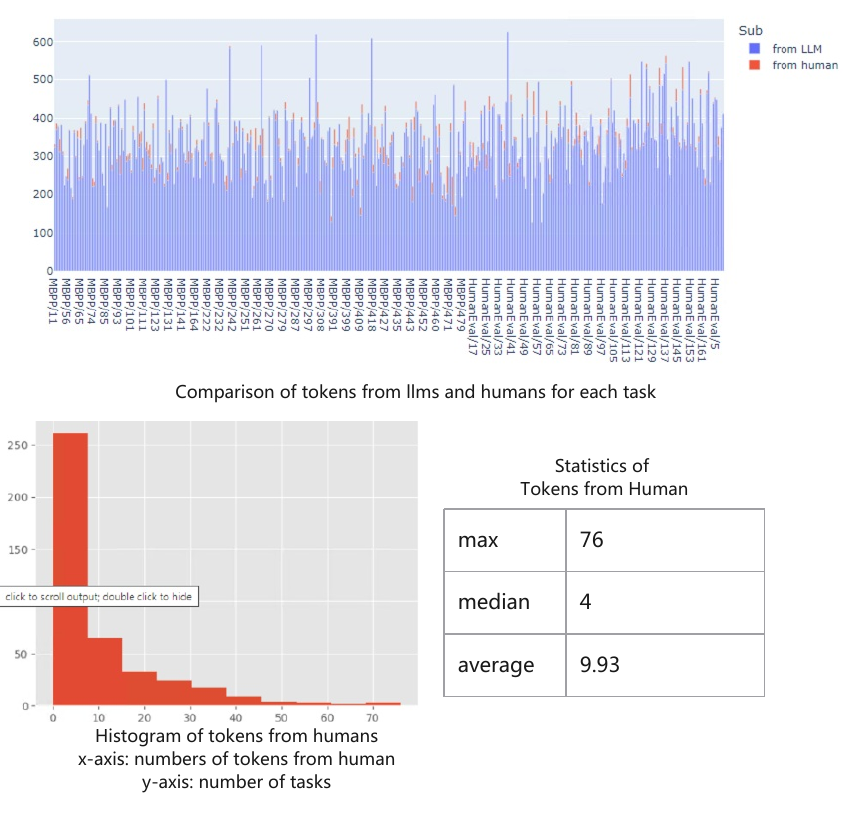}
    \caption{Statistics of Human Labor Savings}
    \label{fig:labor_savings}
\end{figure}

We evaluate the savings of the human labour costs by calculating how many tokens in the final refined requirement specifications are from humans. The statistics are shown in Figure \ref{fig:labor_savings}. From Figure \ref{fig:labor_savings}, we can see that tokens from human users take only a tiny proportion of the refined specifications. To boost the code generation performance, the users need to review the text, delete anything they do not like, and input, on average, ten tokens due to the help of ChatCoder.

\subsection{Relevance and Completeness}

\begin{figure}
    \centering
    \includegraphics[scale=0.4]{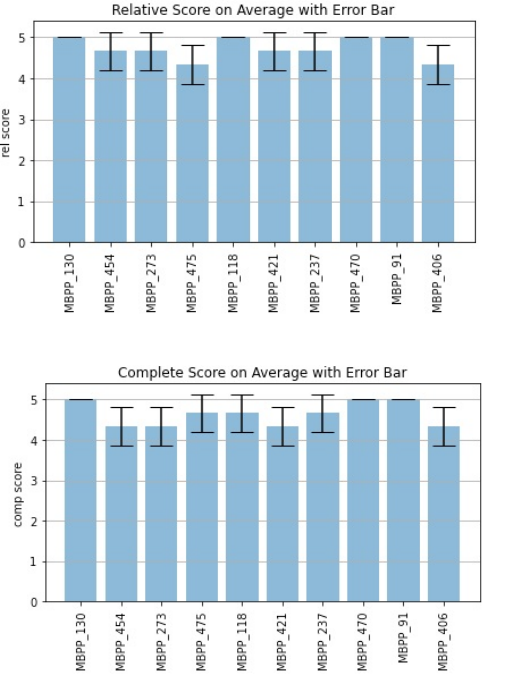}
    \caption{Human Evaluation Score}
    \label{fig:human_eval}
\end{figure}

We need to evaluate whether the improvement is due to ChatCoder's refined requirements and whether the users think ChatCoder's refined requirement specifications fulfil their needs well. Thus we invited three people outside the research group to give scores on ten randomly selected ChatCoder's refined requirements about the `relevance' and `completeness'. The results are depicted in Figure \ref{fig:human_eval}. We ask the testers to compare the requirements before and after refinement and the code generated before and after the requirement refinement. Then we ask them to give a score (1-5) to judge whether the refinement relates to the improvement of the generated code (The real score, 1 for unrelated and 5 for directly related). Besides, we ask them to give a score (1-5) to judge whether the refinement makes them clearer about the user's requirements (The comp score, 1 for getting confused and 5 for getting clear). We calculate the average scores with error bars and have the results depicted in Figure \ref{fig:human_eval}.

Through Figure \ref{fig:human_eval}, we find that all testers agree that the refined requirements help the large language model generate better code and help themselves better understand the requirements. However, compared with the real score, the confidence that people get clearer about the problems is slightly weaker. That is because people judge the quality of the code partially based on the execution test results. However, execution tests are not perfect. The program passing certain test cases may not really fulfil the user's requirements. So ChatCoder still needs to be improved to refine the requirements better to fulfil the user's true needs.

\subsection{Threats to Validations}
There are a few threats to our methods
\begin{itemize}
    \item 1) The user quality. The reviews and edits are performed by the volunteer professional programmers in our research group. So they deeply understand large language models, programming languages and algorithms. However, it can not be guaranteed that every user of large language models is as good at these things as our researchers. So finding some way to test ChatCoder for ordinary users of large language models is on our future work list.
    \item 2) The dataset. We use the datasets, HumanEval and Sanitized-MBPP in this paper to align the other research in this field. However, there is a flaw: these datasets do not really come from 'our requirements' and are too simple compared with real-world programs. One reasonable but difficult-to-realize solution is to recruit a group of full-time programming workers to evaluate the effect of ChatCoder in their real-world job. Finding a more applicable way of evaluating ChatCoder is one of our future work.
    \item 3) The length. The refined requirement specifications obtained by ChatCoder are a bit long. The long text brings two problems. First, it is a heavy burden for the human user to review and edit, leading to a high opportunity to make mistakes. Second, we find through our observation that the current large language models have difficulty in coping with long text: they can ignore the logic dependency of two distanced terms. One of our future works is compressing the refined requirement specifications and preserving all the necessary information. 
    
\end{itemize}

\section{Conclusion}
We propose ChatCoder, an effective method to improve large language models' code generation performances by requirement refinement via chat. We design a two-round dialogue framework to guide the large language model, refine the original requirements through five angles, and go deeper. Then we ask the human users to review and edit the generated refined requirement specifications. We apply ChatCoder to the famous large language models: gpt-4 and gpt-3.5-turbo and prove that ChatCoder improves their code generation ability by a large margin. Besides, we prove that ChatCoder is an efficient way of communicating with the LLMs for requirements refinement, and human intervention is needed in requirements refinement.
\bibliographystyle{ACM-Reference-Format}
\bibliography{sample-base}

\end{document}